\documentclass[journal=jacsat,manuscript=article]{achemso}

\usepackage[version=3]{mhchem} 
\usepackage{xcolor}
\usepackage{kotex}

\usepackage{soul}

\author{Kwang Hyun Cho}
\affiliation[KIAS]{Korea Institute for Advanced Study, Seoul 02455, South Korea}
\author{Seogjoo J. Jang}
\affiliation[CUNY1]{Department of Chemistry and Biochemistry, Queens College, City University of New York, Queens, New York 11367, United States}
\alsoaffiliation[CUNY2]{Ph.D. Programs in Chemistry and Physics, Graduate Center of the City University of New York, New York, New York 10016, United States}
\alsoaffiliation[KIAS]{Korea Institute for Advanced Study, Seoul 02455, South Korea}
\email{seogjoo.jang@qc.cuny.edu}
\author{Young Min Rhee}
\affiliation[KAIST]{Department of Chemistry, Korea Advanced Institute of Science and Technology (KAIST), Daejeon 34141, South Korea}
\email{ymrhee@kaist.ac.kr}
\title[An \textsf{achemso} demo]
  {Minimization of disorder as a key design principle for natural sizes of light harvesting 2 complexes}

\abbreviations{}

\begin{document}

\begin{tocentry}
\centering
\includegraphics[width=0.85\textwidth]{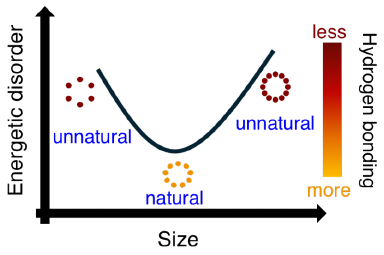}
\end{tocentry}

\begin{abstract} 
The light harvesting 2 (LH2) complex of purple bacteria has excellent energy conversion efficiency. Clarifying the design principle behind such efficiency at the atomistic level is crucial for understanding its structure-function relationship, and can be utilized for the design of artificial light harvesting systems.  
To this end, we conducted comprehensive computational investigation of the dynamical and statistical nature of electronic excited states of pigment molecules in a natural LH2 complex with 9-fold symmetry and its two non-natural {\it in silico} analogues with 6- and 12-fold symmetries. To ensure reliable and efficient all-atomistic molecular dynamics simulations, we combined a well established interpolation approach for the construction of the potential energy surface with a neural network machine learning approach.  Outcomes of these calculations clarify that non-natural forms of LH2-type complexes have significantly larger quasistatic disorder than those for the natural one. 
In addition, non-natural systems have more disruptions of the hydrogen bonding, underscoring its crucial role for reducing the disorder.  On the other hand, local environmental dynamics are relatively insensitive to the structural changes although there is moderate enhancement in the anharmonic or interatomic components for the synthetic ones.
These findings based on all-atomistic simulations provide direct computational evidence that the structure and sizes of natural LH2 complexes are designed to minimize the energetic disorder. We analyze quantitative implications of these for the energy transferring capability of the LH2 complex.

\end{abstract}
Photosynthesis begins with the conversion of light energy into electronic excitation energy and ensuing transport of the molecular energy to reaction centers, a process carried out by various antenna or light harvesting complexes (LHCs).~\cite{Mirkovic2017,Curutchet2017,Jang2018}
These LHCs in general have remarkable quantum efficiency, but molecular-level understanding of how they attain such ability still remains challenging despite decades of research.~\cite{Tani2022,Varvelo2023,Runeson2022,Wu2022,Sohang2022,Duan2020,Saito2019,Krisnanda2018,Samuel2018,Scholes2017,Lohner2015,Mattioni2021}
For example, an interesting mechanism that has been proposed and demonstrated by many researchers is the so called environment-assisted transport,~\cite{Jang2018,rebentrost-njp11} according to which optimal level of energetic fluctuations caused by environments are believed to facilitate favorable excitation energy transfer pathways. 
However, atomistic-level understanding of how the antenna complexes achieve such optimized disorder and fluctuations remains unsatisfactory, despite some progress.
Clarifying this relationship is also the first step for genuine biomimetic engineering of successful synthetic LHCs.

One of the primary sources of environmental fluctuations in light harvesting processes is the nuclear dynamics of molecules, which induce perturbations in the electronic states, modulate exciton dynamics, and play other important energetic and dynamic roles.~\cite{Maly2018,Kim2022,Kim2021}
Understanding how details of such molecular dynamics, in correlation with specific molecular architectures, dictate the fluctuations and the disorder of LHCs is a fundamental challenge in elucidating molecular-level design principles for efficient energy transfer systems. 
To this end, carefully designed atomistic level simulations can provide valuable insight into primary molecular-level factors influencing the dynamics and disorder of electronic states. 
 
Light harvesting 2 (LH2) complex is one of the most widely studied antenna complexes and has been the subject of extensive studies to uncover design principles for efficient exciton transfer.~\cite{Jang2018,jang-jpcb105,jang-jpcb111,jang-jpcb115,Jang2015,jang-jpcl9,Kennis1997,Smyth2015,Janosi2006,cleary-pnas110,Lohner2015,Saga2018,cupellini-pr156,cardoso2019}
However, there are still prominent experiments reflecting different views on various issues, including the role of nuclear dynamics.~\cite{Elad2012,Yeh2012,Chin2013,Richard2013,Duan2017}
An earlier work~\cite{Jang2015} addressed this issue to some extent but involved unsatisfactory models and unconfirmed assumptions.
We here provide results of  more direct and unambiguous computational investigation, which is almost free of such assumptions and offers a more reliable information on how structural modulation influences the disorder and local environmental dynamics.
To be more specific, we construct accurate quantum mechanical system-bath models for a natural LH2 complex with a 9-fold symmetry and two non-natural {\it in silico} models for LH2-like complexes  with 6- and 12-fold symmetries. We then conduct comprehensive computational studies to understand how the structural environment of the LH2 complex modulates its energetic fluctuations.
Outcomes of this work significantly strengthen the major conclusion of  the earlier work~\cite{Jang2015} while offering intriguing new details  that help understand the design principles of LH2 at atomistic level.  
In particular, our new results reinforce the view that hydrogen bonding (HB) plays an important role for the modulation of the disorder, an issue that has not been clearly resolved despite recent investigations.\cite{Montemayor2018,cupellini-pr156,qian2022,Timpmann2024}  In addition, methods employed in this work can also be extended to understanding similar design principles for other types of LHCs.

Computational investigation of design principles at atomistic level may appear to be straightforward and involve only standard molecular dynamics (MD) simulations, from which all nuclear degrees of freedom (DOFs) and their effects can be analyzed.
However, for processes involving electronic excited states, development of consistent and accurate enough potential energy surfaces (PESs) remains challenging for naturally non-existent {\it in silico} complexes as well as naturally existing ones.
Pre-defined force fields cannot ensure such consistency across different LH2 architectures, while quantum mechanics / molecular mechanics (QM/MM) approaches are computationally expensive for long and extensive sampling.~\cite{Maity2020,lin2007}
The major computational breakthrough in methodology we accomplish in this work is to address  this very issue, augmenting a well established interpolation approach called the interpolation mechanics/molecular mechanics (IM/MM) \cite{Ischtwan1994,Thompson1999,collins2002,Park2016,Kim2018,Kim2024,Park2012} method with a neural network machine learning approach.
The superiority of this approach to the previously adopted computational protocol is validated through test calculations (see below and the Supporting Information (SI)).
Outcomes of the resulting all-atomistic MD simulations enable systematic comparisons of the dynamic fluctuations and the statistics of quasistatic disorder for LH2-like complexes, which help elucidating the relationship between the structure/size and the parameters detrimental to the functionality of energy transport.

With recent advances in computational techniques, it has been shown that MD simulation is effective in capturing environmental effects from various perspectives, such as their molecular origin or the time scale.~\cite{Olbrich2011, Loco2018,Rancova2014,Kim2018,Maity2020,Cupellini2018,Maity2021,Samuel2018,Maria2018,Cho2024}
Still, a major challenge in conducting MD simulations of complex systems is the construction of a reliable potential energy surface (PES) governing  the dynamics.
Numerous approaches for delivering PESs have been developed, ranging from empirical functional forms to on-the-fly electronic structure calculations, for which trade-off between computational cost and accuracy is a major determining factor.~\cite{Shen2016,Scott2018,Van2013}
In earlier studies, it was demonstrated that the IM/MM method offers an effective balance between computational efficiency and accuracy in describing various biological processes.~\cite{Kim2024,Kim2018,Park2016}
This is achieved by evaluating potential energies through interpolation of local information pre-computed at reasonably dense set of data points.  A summary of this IM/MM approach and its effectiveness are provided in the Supporting Information (SI).

The effectiveness of IM/MM relies on properly selecting molecular geometries for the precomputation, as they are crucial for ensuring interpolation accuracy and enhancing both computational efficiency and robustness. 
We have previously developed a data-driven approach that employs a genetic algorithm for the filtering, which demonstrated promising results.~\cite{Cho2019}
However, the method requires additional quantum chemistry (QC) calculations, namely, full electronic structure calculations still at excessively many geometries, which limits the feasibility of full automation. In this work, we employ an alternative machine learning (ML) approach that utilizes neural networks that are properly trained so as to filter out data points solely based on the geometric information of the molecule.
Details of this method, along with the description of the overall procedure of the calculation and simulation, are provided below.
\noindent

The focus of our study in this work is the LH2 complex of a purple bacterium called \textit{Rhodoblastus (Rbs.) acidophilus}.~\cite{cogdell-qrb39}, which was formerly known as \textit{Rhodopseudomonas acidophila}. 
It  consists of 27 bacteriochlorophyll (BChl) pigments and surrounding proteins, arranged to have a 9-fold rotational symmetry.
BChls are organized into two distinctive units called  B800 and B850, which respectively absorb light in 800 nm and 850 nm wavelength regions. 
The B850 unit consists of 18 BChls, referred to as BChl-$\alpha$ and -$\beta$, which form two concentric rings with similar radii. The B800 unit constitutes another ring of a larger size with a vertical displacement from the plane of B850 by about 2 nm.  We denote the 9 BChls constituting this unit as BChl-$\gamma$.

Database for the pigment BChl was already available from an IM/MM model for the Fenna-Matthews-Olson (FMO) complex.~\cite{Kim2016}
Thus, the BChl structures in the LH2 database could be prepared from those in the FMO dataset by adjusting the geometries with a displacement vector, $\Delta \textbf{X}=\textbf{X}_{\textrm{LH2}}-\textbf{X}_{\textrm{FMO}}$, where $\textbf{X}_{C}$ represents the optimized BChl geometries within the complex $C$, representing LH2 or FMO.~\cite{Cho2019}
This procedure makes it possible to construct the database of a new complex much faster than with a previously established approach called GROW scheme,~\cite{Kim2016} in which the database is iteratively improved by adding more data points from preliminary simulations.
However, some of the displaced geometries may cause numerical issues for interpolation.
For example, some geometries may bear negative Hessian eigenvalues, which can present unphysically low-energy trapping regions in some part of the interpolated PES.
Although the issue can be resolved through additional application of the GROW scheme, it is more desirable to filter out problematic data points from the dataset in a more systematic manner.
Therefore, additional fine tuning is still required, or an inaccurate energy prediction may destabilize MD simulations by pulling molecules into unphysical regions, which
may even render the GROW scheme inapplicable.

We first optimized the BChl pigment in the native 9-fold LH2 complex using quantum mechanics/molecular mechanics (QM/MM) energy minimization using combination of GROMACS package~\cite{Pronk2013} with Q-Chem 5.0.~\cite{Epifanovsky2021}
Using the optimized structure, we generated a primitive dataset for the LH2 complex by applying parallel displacements to geometries from the FMO dataset and performing QC calculations.\cite{Cho2019}
For these QC calculations, we employed the density functional theory
(DFT) method with the B3LYP functional and a relatively small basis (3-21G) to keep the computational cost manageable.
Then, we ran 10 distinct IM/MM simulations for 2 ps with in-house modified version of GROMACS 4.5~\cite{Kim2018,Kim2016,Cho2019,Cho2024} using this primitive surface. From this trajectory, the geometries were sampled every 20 fs. The next key step was to assess the quality of each data point in the primitive dataset.
We assumed that data points that yield inaccurate potential energies for the sampled geometries should be considered unreliable, as they provide an inaccurate description of the energy landscape, at least in certain regions.
Indeed, removing such geometries practically ensures stable simulations.

To quantify issues as described above, we calculated the mean squared error (MSE) between the potential energies predicted from each data point using a second-order Taylor expansion and the corresponding reference QC energies as follows.
\begin{align}
\label{eqn:mse}
E_i=\frac{1}{M}\sum^{M}_{\alpha=1} \left(V_i(\textbf{Z}_\alpha)-V_\alpha\right)^2
\end{align}
Here, $\alpha$ and $M$ denote the index and the total number of sampled geometries, while $i$ denotes the index in the primitive dataset.
$V_i(\textbf{Z}_\alpha)$ refers to the potential energy of a sampled geometry, namely, $\textbf{Z}_\alpha$ estimated using the $i$-th data point, $V_i(\textbf{Z}_\alpha)=E_i + \textbf{D}^\textrm{T}_i\cdot \textbf{g}_i+\frac{1}{2}\textbf{D}^\textrm{T}_i\cdot \textbf{h}_i\cdot\textbf{D}_i$ with $\textbf{D}_i=\textbf{Z}_\alpha-\textbf{Z}_i$.
$V_\alpha$ is the actual QC energy of the sampled geometry $\textbf{Z}_\alpha$ calculated with the B3LYP functional and the 3-21G basis set. 
We used this error to evaluate each displaced geometry $i$ and selected those with the smallest values. Through this process, we selected 400 data points of displaced geometries with smallest errors, which served as the dataset for the actual IM PES. At this stage, QC calculations were re-performed with a larger basis set on these selected configurations, namely at the B3LYP/6-31G(d,p) level, to ensure the accuracy of the PES. We will later demonstrate the accuracy of this PES, showing that the MSE is a reliable indicator for selecting appropriate geometries.

Even with a small basis set, performing QC calculations for many geometries for the preliminary selection stage is still a demanding task. To avoid repeating this protocol for other LH2 systems with different symmetries, we trained a fully connected neural network (FNN)~\cite{rumelhart1986} to predict the squared error $E_i$ of any given data point in the primitive dataset based on its molecular geometry.
The input features to the neural network were the molecular geometries, in internal coordinates in the Z-matrix form.
These coordinates were expressed as deviations from the corresponding values of the optimized reference structure.

The network contained three hidden layers with 300, 450 and 150 neurons, respectively, each using the ReLU activation function,~\cite{nair2010} which had approximately 240,000 trainable parameters.~\cite{nair2010}
Hyperparameters, such as the number of hidden layers and the learning rate, were tuned empirically to optimize performance.
For each type of the BChl, an independent FNN was trained on a full set of geometries in the primitive dataset, which was randomly split into training and test sets with a ratio of 80:20.
The model parameters were optimized using the Adam optimizer~\cite{kingma2017} with the MSE as the loss function. 
Model performance was evaluated on a test set and a scatter plot comparing the predicted and reference values is provided in Figure S1, demonstrating the reliability of the model.
Once trained, the network was used to predict the errors for the BChl geometries in the LH2 complexes with other symmetries.
Based on these predictions, we could select 400 geometries with the lowest expected errors as the most reliable data points. For these data, electronic structure calculations were conducted at the same B3LYP/6-31G(d,p) level to construct the final IM PES.
Evaluation of the excited state PES was made by conducting linear response time-dependent (TD)-DFT~\cite{casida2012} calculations for the same data set, enabling straightforward access to excited state quantities, such as excitation energies, which has been difficult to realize earlier.~\cite{Jang2015}

The interaction between the IM and the MM regions was described using Coulomb interactions within a fixed charge model. 
For each complex, the ground state partial charges of the pigments were optimized using QM/MM calculations and the restrained electrostatic (RESP) method (see Table S1 in the SI for the values of fixed charges used in our simulation).~\cite{bayly1993,Kim2018}
To better reproduce the gap energy fluctuations, the excited state partial charges were optimized to match QM/MM results, rather than being determined with RESP fitting.~\cite{Kim2018}
This strategy provides a reasonable description of the gap energy fluctuations, with their magnitudes consistent with those obtained from the electrostatically embedding QM/MM approach (Figure S2).

\begin{figure}[t!]
\centering
\includegraphics{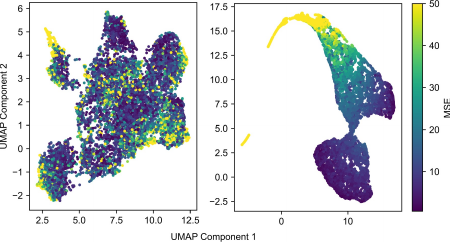}
  \caption{Distribution of molecular geometries in the primitive dataset for BChl $\alpha$. The geometries were visualized by projecting its internal coordinate, namely the input for the neural network (left) and the latent space represented by the last hidden layer (right) with UMAP. The color of each point represents MSE of each geometry calculated using eq \ref{eqn:mse}.}
  \label{fig:umap}
\end{figure}

Before performing all-atomistic simulations with the obtained PESs, we examined how the neural network approach selects appropriate data points for the IM dataset.
To this end, we visualized the distribution of molecular geometries in the primitive dataset using uniform manifold approximation and projection (UMAP) as a dimensionality-reduction technique for effective visualization.~\cite{mcinnes2020}
Initially, the molecular geometries were represented by their internal coordinates vector, which served as the input descriptor for the neural network.
UMAP was then applied to project these descriptors into two dimensions for visualization purposes (Figure~\ref{fig:umap}, left panel).
As shown, the raw descriptors displayed no notable structure with respect to the MSE, making it difficult to estimate the quality of geometries based on their molecular descriptions.
In contrast, after passing through the first two layers of our FNN model, the molecular geometries were encoded into a 150-dimensional vector.
When visualized using UMAP (Figure~\ref{fig:umap}, right panel), these transformed vectors exhibited a much more organized structure, where the MSE values were smoothly distributed.
Notably, data points with relatively high MSE values formed a distinct tail-like pattern, effectively identifying regions of configuration space less suitable for PES dataset.
This result suggests that the neural network effectively encodes molecular geometries, extracting task-relevant descriptors that are highly informative for the selection of optimal configurations for PES construction.

\begin{figure}[t!]
\centering
\includegraphics{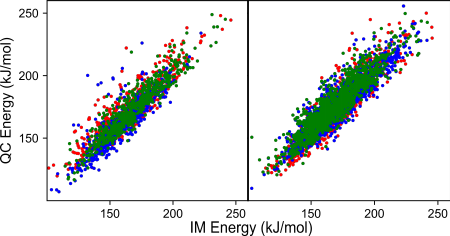}
  \caption{Correlation between the estimated potential energy and the reference potential energy. The left panel represents the 6-fold symmetric LH2, while the right panel represents the 12-fold symmetry. The color of the dots indicates the type of BChl: BChl $\alpha$ is shown in red, $\beta$ in blue, and $\gamma$ in green.}
  \label{fig:imcorr}
\end{figure}

The quality of the interpolated PESs for the natural LH2 complex and other synthetic complexes with different symmetries were evaluated and assessed as described below.
For the evaluation, we conducted 500 ps all-atomistic simulations employing the IM/MM approach at constant temperature of 300 K with the velocity rescaling thermostat.~\cite{Bussi2007}
Out of all the trajectory snapshots, we extracted 50 configurations for each BChl at 10 ps intervals. 
Thus, for a system with $n$-fold symmetry, a total of $50\times n$ geometries were collected for each type of BChl. 
The potential energies computed using the interpolated scheme were then compared with the reference energies of DFT calculations.
Figure~\ref{fig:imcorr} provides the correlation data between the IM energies and the reference full electronic structure calculation energies for each BChl in 6-fold and 12-fold symmetry LH2-type complexes, with root-mean-squared (RMS) errors of $\sim$0.1 eV.
This excellent correlation validates the accuracy of the IM PES, and demonstrates  that the neural network approach effectively sampled proper geometries for the interpolation dataset. 
That is, the neural network approach can reliably identify geometries that are likely to cause potential problems, for example, interpolation instability, and thereby serves as a data selection algorithm that enables transfer of the dataset to closely related systems.

This way, we could minimize the computational cost of constructing IM PESs toward systematically investigating for synthetic LH2-type complexes with different symmetries, while maintaining consistency in treating core pigments.

\begin{figure}[t!]
\centering
\centering
\includegraphics[width=11.4cm]{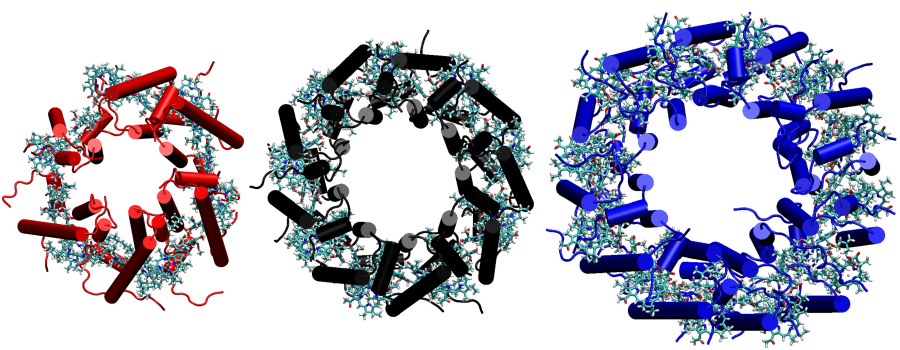}
  \caption{Equilibrated complexes with 6-fold (red), 9-fold (black) and 12-fold (blue) symmetries. BChls in the complexes are also visualized.}
  \label{fig:complex}
\end{figure}

Utilizing the IM PESs, we conducted all-atomistic simulations to analyze energetic variations of each symmetric LH2 complex.
Starting from the energy minimized structure, as described in SI, the system was then relaxed through 1 ns of NPT equilibration at a constant temperature of 300 K and a pressure of 1 atm, employing the velocity rescaling thermostat~\cite{Bussi2007} and the Parrinello-Rahman barostat.~\cite{Parrinello1981}
Figure~\ref{fig:complex} depicts the snapshots of the three complexes.
Subsequently, the system was heated at 500 K for 2 ns to ensure diverse sampling of the complex, using the same thermostat.
During the last 1 ns of the high temperature sampling, 10 frames were recorded at intervals of 100 ps, which were used as initial configurations for the subsequent process.
Each of the 10 configurations was then re-equilibrated at 300 K with NVT simulation for another 1 ns. 
Following this, each simulation was further extended for another 100 ps with NVT, during which geometries and excitation energies of BChls were recorded every 5 fs utilizing the excited PES.
Due to the $n$-fold symmetry of LH2, with the 10 trajectories, we could obtain effectively $n\times10$ trajectories of BChls.
A simplified overview of the protocol is presented in Figure S3.

\begin{figure}[t!]
\centering
\includegraphics[width=11.4cm]{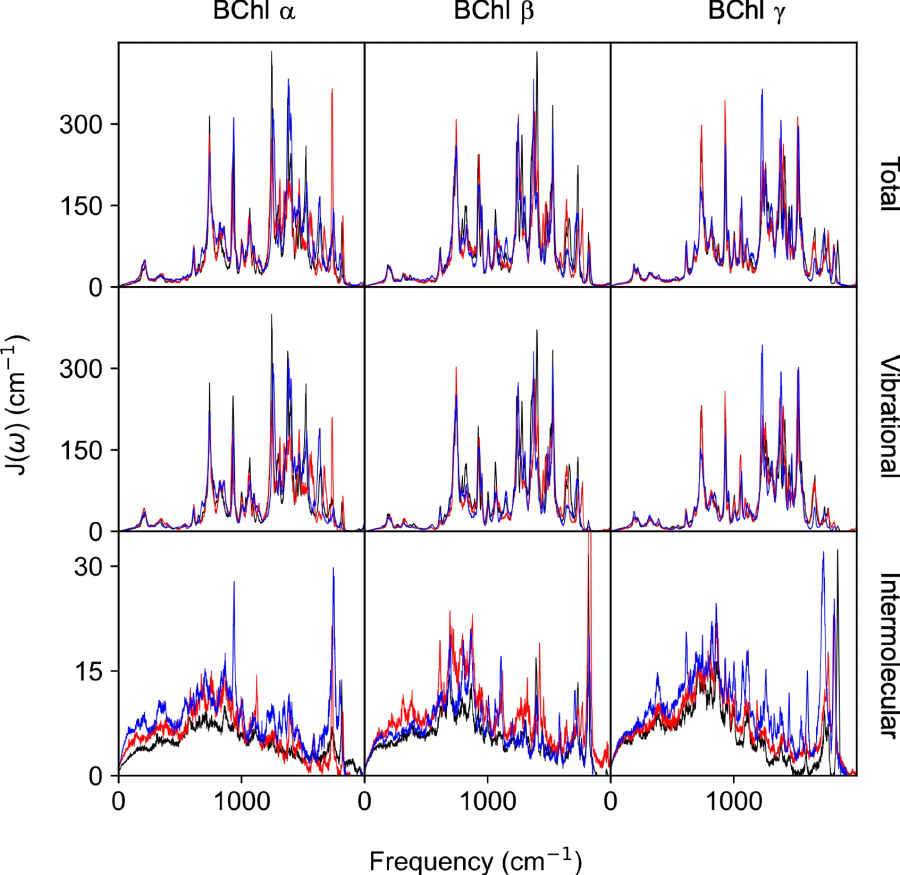}
  \caption{Spectral densities of BChl $\alpha$ (left column), BChl $\beta$ (middle column), and BChl $\gamma$ (right column)  are shown. Each row represents a specific component of the spectral density: total spectral density (top row), vibrational spectral density (middle row), and intermolecular interaction (bottom row). The line color indicates the symmetry of LH2 complex, red for 6-fold, blue for 12-fold and black for 9-fold symmetry.}
  \label{fig:spectral_density}
\end{figure}

First, let us focus on the dynamic fluctuations in excitation energy captured within the 100 ps simulations.
As described already, in all trajectories, we recorded time series of the excitation energies at every 5 fs as $\Delta E_{\nu,i}(t)$, defined relative to the average value. Here, $i$ is the trajectory index while $\nu$ refers to the site index of BChl.
From this, we computed the autocorrelation function up to 10 ps as below:
\begin{align}
    C(t)=\frac{1}{n\cdot N}\sum^{n}_{\nu=1}\sum^{N}_{i=1}\left\langle\Delta E_{\nu,i}(\tau)\cdot \Delta E_{\nu,i}(t+\tau)\right\rangle_\tau .
\end{align}
After padding this up to 20 ps by zero, we then calculated the spectral density employing the following expression: 
\begin{align}
    J(\omega)=\frac{2}{\pi\hbar}\frac{\beta\hbar\omega}{2}\int^\infty_0d \tau C(\tau)\cos(\omega\tau) .
\end{align}
We found the simulation duration of 100 ps to be sufficient for capturing representative fluctuations, as the calculated spectral densities reached statistical convergence.
Indeed, we found that extending the trajectories up to 1 ns yielded visually identical results.
Spectral densities of BChls within the three complexes shown in Figure~\ref{fig:complex} were compared to quantify the environmental perturbations associated with different symmetries. 
The first row of Figure~\ref{fig:spectral_density} shows the calculated spectral densities of BChls, where no notable differences are apparent at a first glance, despite the underlying symmetry changes.
Nevertheless, decomposition of the total environmental fluctuations revealed more subtle, symmetry-dependent characteristics that were otherwise obscured in the total spectral densities.
In fact, the excitation energy can be decomposed into an intra-pigment component and a residual fluctuation, which is commonly attributed to the interaction with surrounding media, such as protein environment and solvent.
The total spectral density can consequently be decomposed into: (1) the portion from pigment vibrations and (2) a residual part,~\cite{Cho2024,Kundu2022,Schulze2016} which includes contributions of intermolecular interactions.
The second and third rows of Figure~\ref{fig:spectral_density} show that the intra-pigment vibrational contributions give rise to high-frequency peaks, whereas the remaining part exhibits a relatively featureless broad phonon sideband.
Now, it becomes evident that vibrational spectral densities are largely unaffected by changes in the symmetry of the LH2 complex, in terms of peak positions, relative intensities, and spectral broadening. 
This indicates that the pigment vibrations and their couplings to the excitation energies are robust against changes in protein symmetry.
On the other hand, notable differences are observed in the residual part of the bath spectral density, particularly for BChls in the B850 unit.
Compared to the other symmetries, the spectral densities of the pigment in 9-fold symmetry is significantly reduced, particularly in the low-frequency region.
The reduced spectral densities indicate that fluctuations in the excitation energies are suppressed, which is also reflected in the reorganization energy of the residual part of the bath spectral density, as shown in Table~\ref{tbl:reorg}.
These results suggest that the BChl molecules exhibit a constrained positional fluctuation with respect to the surrounding environments.
That is, the BChls in the B850 unit are more tightly packed and exhibit more suppressed motion compared to those in the non-natural symmetric LH2 complexes, resulting in reduced dynamic intermolecular interactions.

\begin{table*}[t!]
\centering
  \caption{Reorganization energy (in cm$^{-1}$).}
  \label{tbl:reorg}
  \begin{tabular}{c|ccc|ccc|ccc}
    \hline
    Symmetry & \multicolumn{3}{c|}{6-fold} & \multicolumn{3}{c|}{9-fold} & \multicolumn{3}{c}{12-fold} \\
    \hline
    BChl type  & $\alpha$ & $\beta$ & $\gamma$ & $\alpha$ & $\beta$ & $\gamma$ & $\alpha$ & $\beta$ & $\gamma$ \\
    \hline
    Total                      & 134 &  133 & 131 &124&  137  & 130 & 151 & 137 & 139 \\
    Pigment vibration           & 102 & 98  & 97& 103 & 112 & 97 & 109 & 99 & 97  \\
    Intermolecular interactions & 28 & 36 &  34 &19 & 25 & 29 & 29 & 29  & 39\\
    \hline
  \end{tabular}
\end{table*}

Extracted bath spectral densities are useful tools for investigating environmental perturbations. Nonetheless, certain aspects require further attention.
An important consideration is the limited timescale of 100 ps, which excludes slower fluctuations.
Fluctuations arising from certain DOFs, such as reorganization of the solvent or protein movement, are slow enough to persist beyond the nanosecond time scale.~\cite{Kim2018}
When these slower movements occur on a timescale much longer than that of the relevant exciton transfer, their dynamical contributions are insignificant. Instead, they introduce  quasistatic disorder into the system, which is commonly referred to as static disorder.~\cite{Chaillet2020,Tjaart2017,Kulkarni2025}
In LHCs, quasistatically locked protein structures surrounding pigment molecules are sources of the static disorder, leading to non-uniform excitation energies of pigments.
The disorder effectively breaks the symmetry of the electronic system, which is known to modulate the exciton transfer pathway and rates, potentially influencing the robustness of energy transfer.~\cite{Kulkarni2025,Kennis1997,Maly2018}
Despite its importance, quantifying static disorder is nontrivial and has largely relied on phenomenological analyses.~\cite{Vladimir2006,Rancova2014,Renger2002}
Also, understanding the microscopic origin of the disorder, especially from the perspective of atomistic MD simulations remains ambiguous and largely unexplored.


Here, we will discuss the presence and characteristics of the quasistatic disorder that are gleaned from the ensemble of trajectories we generated. 
Although each trajectory spans only a certain timescale, their ensemble collectively samples diverse regions of conformational space, allowing us to infer variations that would otherwise emerge over longer simulations.
For this purpose, we collected the time-averaged excitation energy of each trajectory,
\begin{align}
    \tilde{E}_{\nu,i} = \frac{1}{T}\int^T_{t=0} \Delta E_{\nu,i}(t) dt ,
\end{align}
where $T$ is the time scale of each trajectory, set to 100 ps. 
We emphasize that 100 ps is not sufficiently long to sample all environmental perturbations, and the values of $\tilde{E}_{\nu,i}$ will naturally fluctuate over the site index $\nu$ and the trajectory index $i$. 
Indeed, $\sigma_{\rm{total}}$ defined as the standard deviation in the values of $\tilde{E}_{\nu,i}$ over all $\nu$ and $i$ does not vanish as shown in Table~\ref{tbl:std}. 
There are at least two sources of fluctuations for $\tilde{E}_{\nu,i}$: the $i$-dependent one that was inherited from the thermal sampling stage over some nanoseconds, and the $\nu$-dependent one that reflects how much distortion each LH2 has from its ideal $n$-fold symmetry. 
Before further discussing on how to distinguish these two aspects, from Table~\ref{tbl:std} we can notice that the $\sigma_{\rm{total}}$ values for BChls in the B850 ring are significantly smaller in the 9-fold symmetry compared to the non-natural symmetries.
This hints us that the quasistatic disorder in the B850 unit, with timescales longer than 100 ps, is a minimum for  the 9-fold symmetry of the LH2 complex. Interestingly,
for BChl $\gamma$, $\sigma_{\rm{total}}$ values are comparable over different symmetries with slight increases with increasing $n$.

Let us now introduce two additional statistical averages: (1) averaging the excitation energies of all BChls within each trajectory (trajectory-indexed) and (2) averaging the excitation energies of each site of BChl across trajectories (site-indexed).~\cite{Agarwal2001}
These are defined as:
\begin{align}
    \label{eqn:average}
    \tilde{E}^{\rm{traj}}_{i}=\frac{1}{N_{\rm{site}}} \sum^{N_{\rm{site}}}_{\nu=1} \tilde{E}_{\nu,i} \nonumber \\
    \tilde{E}^{\rm{site}}_{\nu}=\frac{1}{N_{\rm{traj}}} \sum^{N_{\rm{traj}}}_{i=1} \tilde{E}_{\nu,i}
\end{align}
The former characterizes trajectory-dependent fluctuations by averaging the excitation energy across all sites within a single trajectory, while the latter captures site-dependent variations by averaging the excitation energy of each BChl site over different trajectories.
Corresponding standard deviations are respectively referred to as $\sigma_{\rm{traj}}$ and $\sigma_{\rm{site}}$:
\begin{align}
    \sigma_{\rm{traj}}&=\sqrt{\frac{1}{N_{\rm{traj}}-1}\sum_{i=1}^{N_{\rm{traj}}} (\delta \tilde{E}^{\rm{traj}}_i)^2}\\
    \sigma_{\rm{site}}&=\sqrt{\frac{1}{N_{\rm{site}}-1}\sum_{\nu=1}^{N_{\rm{site}}} (\delta \tilde{E}^{\rm{site}}_\nu)^2}
\end{align}
where $(\delta \cdots)^2$ implies variance around the average of the distribution. 
$\sigma_{\rm{traj}}$ includes disorder arising from different initial conditions of configurations, while $\sigma_{\rm{site}}$ reflects gap energy deviations regarding the pigment site arising from differences in local protein environments.
 The assumption that all individual trajectories are independent of each other and that BChls at different sites are all independent would mean that the two standard deviations are identical.
However, we found that the two standard deviations differ significantly.
A general trend is that $\sigma_{\rm{site}}$ is substantially larger than $\sigma_{\rm{traj}}$, as seen in Table~\ref{tbl:std}.

\begin{figure}[t!]
\centering
\includegraphics[width=10.4cm]{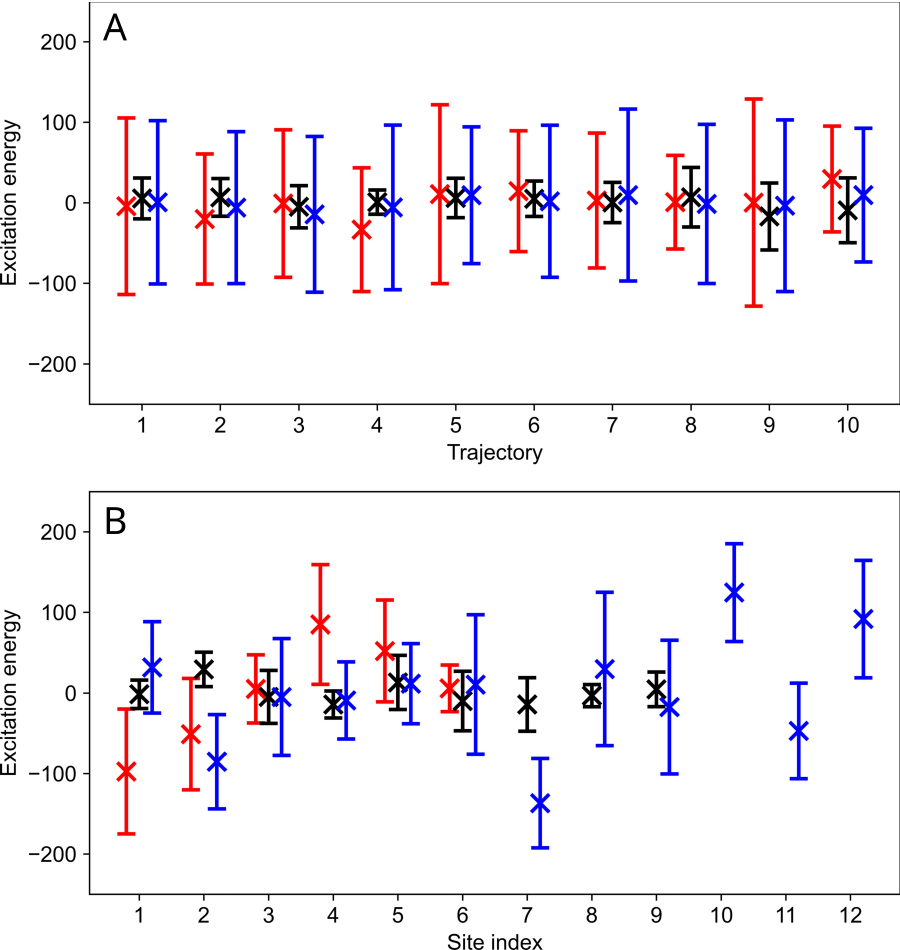}
  \caption{Averages (cross marks) and standard deviations (indicated by two horizontal bars)  of the trajectory-indexed and site-indexed average excitation energies of BChl $\alpha$ for the three complexes. Black color represents the natural complex with 9-fold symmetry, red the 6-fold synthetic one, and blue the 12-fold synthetic one.  Panel A shows the data based on the trajectory-indexed average energy, $\tilde{E}_i^{\mathbf{traj}}$, for different trajectories, whereas panel B shows the data based on the site-indexed average energy, $\tilde{E}_{\nu}^{\mathbf{site}}$. Therefore, in A, the length of each bar represents how much difference a BChl $\alpha$ site can display depending on its location, whereas in B, the length of each bar exhibits how different each site can be in different trajectories.}
  \label{fig:mean_std}
\end{figure}

To investigate the origin of the discrepancy between the two standard deviations, we visualized the time-averaged excitation energies of BChl $\alpha$, categorized by site and trajectory indices (Figure~\ref{fig:mean_std}).
The result shows the site specific variations persisting across different trajectories.
Namely, certain BChl $\alpha$ pigments consistently exhibit higher or lower excitation energies depending on their position.
For instance, in the 12-fold LH2, the 10th BChl $\alpha$ shows significantly higher excitation energy than the 7th, highlighting significant inhomogeneity of site excitation energies regarding their site.
This observation suggests that, despite the rotational symmetry of the LH2, individual BChl $\alpha$ are embedded in distinct local protein environments that remain locked during the duration of electronically excited states.
Each trajectory was initiated from structures that were sampled at 100 ps intervals, during which the site dependent energetic patterns remained stable.
This indicates that once a specific protein environment is established, transitioning between temporally distinct protein structures may occur on timescales longer than nanoseconds.
It introduces long-lived and site-dependent energetic heterogeneity, which is reflected in an increase in $\sigma_{\rm{site}}$.

Interestingly, in non-natural symmetry cases, we found that for BChl in the B850 ring, $\sigma_{\rm{site}}$ is more than three times greater than $\sigma_{\rm{traj}}$ while the ratio is less than two in its native symmetry.
Recalling that the site correlation reflects the sluggishness of the temporarily formed protein environment, we can infer that with the 9-fold symmetry, heterogeneous protein environments are more effectively compensated during the annealing process.
Thus, we can expect that, not only the magnitude but also the timescale representing quasistatic disorder is effectively reduced in this natural case, providing more homogeneous distribution of pigments.
This finding suggests that the natural symmetry of LH2 plays a central functional role in regulating excitation energy distributions, thus optimizing exciton transfer efficiency.~\cite{Jang2015}
 
Although the number of trajectories used in the above analysis is relatively small, they are representative enough to demonstrate important qualitative trends. With more extensive sampling of trajectories for more initial conditions and longer trajectories, it is expected that quantitative modeling of the disorder and fluctuations will become feasible.

\begin{table*}[t!]
\centering
  \caption{Sample standard deviation in mean excitation energy (in cm$^{-1}$).}
  \label{tbl:std}
  \begin{tabular}{c|ccc|ccc|ccc}
    \hline
    Symmetry & \multicolumn{3}{c|}{6-fold} & \multicolumn{3}{c|}{9-fold} & \multicolumn{3}{c}{12-fold} \\
    \hline
    BChl type  & $\alpha$ & $\beta$ & $\gamma$ & $\alpha$ & $\beta$ & $\gamma$ & $\alpha$ & $\beta$ & $\gamma$ \\
    \hline
    $\sigma_{\rm{total}}$          & 85.11 & 77.25 & 62.28 & 28.74 & 47.53 & 79.91 & 93.85 & 63.07 & 86.43 \\
    $\sigma_{\rm{traj}}$          & 17.48 & 20.78 & 17.15 & 8.05  & 17.54 & 19.18 & 7.92 & 9.99 & 19.54 \\
    $\sigma_{\rm{site}}$          & 66.50 & 58.27 & 50.09 & 14.11 & 29.03 & 67.23 & 70.53 & 45.48 & 67.17 \\
    \hline
  \end{tabular}
\end{table*}

We also discovered possible molecular architectures with the symmetric structure, at least partially, that contribute to optimizing energetic variations in the natural complex.
Previous studies have reported substantial influence of HB on the site energies of light harvesting complexes.~\cite {Montemayor2018,cupellini-pr156,qian2022,Timpmann2024,Zhu2025}. 
In the LH2 complex, the key HBs involve the acetyl groups of BChl $\alpha$ and $\beta$, which form bonds with the amine group of a tryptophan and the hydroxyl group of a tyrosine, referred to as HB-$\alpha$ and HB-$\beta$, respectively.
Additionally, an alternative HB-$\beta'$, formed between the acetyl group of BChl $\beta$ and another tryptophan has been proposed, suggesting a stabilization mechanisms for pigments within the protein scaffold.~\cite{Jang2015}
To evaluate the formation probabilities of these HBs, we determined the statistics of HBs from the MD trajectories using a weighting function used in a previous work.~\cite{Jang2015}
The results are summarized in Table~\ref{tbl:hbond}, which are quantitatively different from previous data \cite{Jang2015} but are qualitatively consistent.
In the natural 9-fold symmetric LH2 complex, both HB-$\alpha$ and HB-$\beta$ are present, serving as anchors that stabilize the pigments within their binding pockets.
In contrast, under altered symmetries, these HBs are frequently disrupted.
This leads to BChls exhibiting greater freedom of motion within their surrounding environment.
Although Figure~\ref{fig:spectral_density} shows that such disruptions have little effects on the vibrational structure of the pigment molecules, they contribute to enhancing fluctuations of intermolecular interactions. Most importantly, disruptions of HB are closely correlated with the increase of disorder.  
This suggests that the primary role of the HB network is not to affect the dynamics but rather the energetics of BChls. 
Notably, our results indicate that static disorder is also likely minimized in the naturally symmetric structure.
This is presumably due to the strong and persistent anchoring of pigments at specific sites through the HB network, which prevents pigment motions within the protein matrix and insulates them from fluctuations in the surrounding protein scaffold.
Thus, our computational results are consistent with recent analyses of cryo-EM data.\cite{qian2022} On the other hand, we do not find any reliable evidence that disruption of HB leads to enhancement of electronic couplings between BChls as suggested from spectroscopic data on bio-engineered mutants without HB.\cite{Timpmann2024}

\begin{table}[t!]
\centering
  \caption{Probabilities of HB status}
  \label{tbl:hbond}
  \begin{tabular}{c|ccc}
    \hline
    HB & 6-fold & 9-fold & 12-fold\\
    \hline
    none            & 0.43& 0.02 & 0.81\\
    $\alpha$        & 0.16& 0.24 & 0.13\\
    $\beta^\prime$ &0.05 & 0 & 0.05\\
    $\alpha$,$\beta^\prime$ &0.01 &0 & 0 \\
    $\beta$ &0.28& 0.09& 0.01\\
    $\alpha$,$\beta$ &0.07&0.60 &0 \\
    $\beta$,$\beta'$ &0& 0.01& 0\\
    $\alpha$,$\beta$,$\beta'$ &0 &0.04 &0 \\
    \hline
  \end{tabular}
\end{table}

\begin{figure}[h!]
\centering
\includegraphics[width=3in]{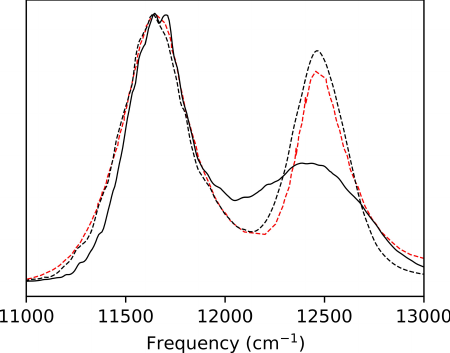}
  \caption{Calculated linear absorption line shape of natural LH2 complex with 9-fold symmetry, in comparison with experimental data (red, dashed). For the former, total standard deviations of excitation energies are  230, 244, 252 cm$^{-1}$ for BChl $\alpha$, $\beta$ and $\gamma$, respectively. Excitation energies are set to $E_\alpha=12150\;\textrm{cm}^{-1}$, $E_\beta=12150\;\textrm{cm}^{-1}$ and $E_\gamma=12550\;\textrm{cm}^{-1}$. Adding additional noise of $30 \textrm{cm}^{-1}$ on BChl $\alpha$ and $\beta$ with scaling $\sigma_\gamma$ by a factor of 0.45 yields better agreement with experiment (black, dashed).}
  \label{fig:abs}
\end{figure}

Finally, we examined how the symmetry-dependent disorder characterized above manifests in actual exciton dynamics.
First, we computed the absorption line shape of the natural LH2 complex with 9-fold symmetry.
A total of 1,000 structures (at intervals of 100 fs) were sampled from a 100 ps IM/MM trajectory.
For each sampled structure, the absorption line shape was computed with modified Redfield theory,~\cite{Vladimir2006,zhang1998,womick2011} with the electronic Hamiltonian given at the sampled structure.
The diagonal elements, corresponding to site excitation energies, were obtained from IM/MM calculations and uniformly shifted so that the spectrum is well aligned with the experimental one.
The off-diagonal couplings were derived for each pair using the transition charges from the electrostatic potential (TrESP) method.~\cite{madjet2006,Kim2016}
The vibrational contribution to the line broadening was realized through the line broadening function derived from the harmonic vibrational component of the spectral density.
The results were then averaged over the ensemble.
The simulated absorption spectrum for the B850 band (Figure~\ref{fig:abs}) is quite close to that of the experimental one even without assuming any additional disorder.
This confirms that our trajectories are effective in representing the majority of the quasistatic disorder. 
Absorption line shapes for the other two complexes with 6- and 12-fold symmetries were also calculated using the same procedure and are provided in the SI (Figure S6), which show broader lineshapes than the natural one.

Figure~\ref{fig:abs} also shows that the agreement between simulated and experimental lineshapes for B850 improves even further when additional random Gaussian disorder with a standard deviation of $30\ {\rm cm^{-1}}$ is introduced in site excitation energies of  BChls $\alpha$ and $\beta$.
Considering the small magnitude of this disorder and lack of correlation of this disorder, we conclude that our MD trajectories serve as reliable representations for quantitative comparison of the effects of quasistatic disorder on exciton dynamics involving B850 units between different complexes.
On the other hand, for the B800 band, our simulated lineshapes are significantly broader than the experimental one.  In fact, the difficulty of reproducing experimental B800 band based on all-atomistic simulations has been reported before.\cite{cupellini-pr156} 
Note that the model employed in Ref. \citenum{cupellini-pr156} accounted for many-body polarization effects of the MM part through polarizable dipoles. Thus, we believe the discrepancy is due to manybody polarization effects that go beyond such level of description.
Within the current MM description of intermolecular interactions, fixed point charges represent the majority of Coulomb interactions. This may produce frustration of charges in the hydrophilic environments of BChl $\gamma$ of the B800 unit that may not exist in actual environments with flexible charge distributions, leading to overestimated influence from the polar environments.
Indeed, better agreement is achieved when fluctuations of BChl $\gamma$ excitation energies from the overall average are rescaled by a factor of 0.45 (see Figure ~\ref{fig:abs}). 
While further computational investigation and more satisfactory analyses remain necessary, these do not have direct implications for inter-LH2 exciton transfer, which is the major focus our work. 
The similar quality of our simulated B850 lineshape for the natural 9-fold LH2 complex in Fig. ~\ref{fig:abs} with that of Ref. \citenum{cupellini-pr156} also confirms that both works reliably account for the majority of the disorder in the B850 part, for which we provide more detailed consideration as detailed below.

\begin{figure}[h!]
\centering
\includegraphics[width=3in]{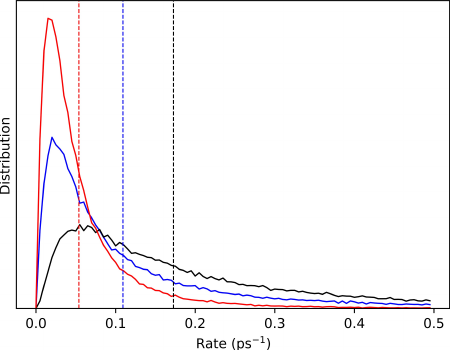}
  \caption{Distribution of inter-LH2 exciton transfer rates for 6-fold (blue), 9-fold (black), and 12-fold (red) symmetries. Mean values are indicated by dashed lines.}
  \label{fig:rate}
\end{figure}

We further computed the inter-LH2 exciton transfer rate for the three complexes.
As excitons reside predominantly in the B850 ring, the inter-LH2 exciton transfer is modeled as that between two B850 rings.
Employing the 100 structures (at an interval of 1 ps) generated from the 100 ps IM/MM trajectory based on which the absorption lineshape in Fig. \ref{fig:abs} was calculated, 10,000 samples of LH2-LH2 pairs were generated by taking every combination of the structures.
For each pair, one LH2 complex was translated parallel to the B850 ring plane such that the shortest distance between two outer $\beta$-rings in the LH2 pair was 2 nm.
The translated complex was then rotated about its $n$-fold symmetry axis by a random angle.
This rotational sampling generated 10 relative orientations per pair, resulting in a total of 100,000 LH2-LH2 configurations.
For each configuration, the electronic Hamiltonian describing the two coupled B850 bands was constructed.
The inter-LH2 exciton transfer rate was calculated for each instance with a generalized master equation for modular exciton density (GME-MED)\cite{jang2014} employing a diagonal approximation in exciton basis, which was demonstrated to provide reasonably accurate results for LH2-LH2 exciton transfer rates.\cite{Jang2018,jang2014}
Detailed rate expressions and calculated time correlation functions, which enter the rate expressions, are provided in SI.
The average and the standard deviation of the distribution of rates for each LH2 complex, shown in Figure~\ref{fig:rate}, are summarized in Table~\ref{tbl:rate}.
The average rate of the 9-fold LH2 complex is close to an experimental observation~\cite{agarwal2002} and exhibits the fastest inter-LH2 exciton transfer rates among the three complexes, consistent with its minimized energetic disorder, reflecting that reducing disorder is related to making inter-complex exciton transfer more efficient.

\begin{table*}[t!]
\centering
  \caption{Exciton transfer rates (in ps$^{-1}$).}
  \label{tbl:rate}
  \begin{tabular}{c|ccc}
    \hline
    Symmetry & {6-fold} & {9-fold} & {12-fold} \\
    \hline
    Average   & 0.109 & 0.173 & 0.054  \\
    Std. Dev.    & 0.155 & 0.193 & 0.076  \\
    \hline
  \end{tabular}
\end{table*}

In conclusion, we have compared the dynamics and statistics of electronic excitations of BChls in a natural LH2 complex that has 9-fold symmetry, with those for two {\it in-silico} synthetic analogues. The analogues were constructed to have different sizes with 6- and 12-fold symmetries while employing identical molecular building blocks.  The potential energy model used in this work was based on the combination of the IM/MM method and an FNN-ML approach, which allowed reliable and efficient evaluation of energies and forces in a manner that was virtually equivalent to QM/MM simulations.
Thus, the present approach addresses major issues of an earlier work~\cite{Jang2015} which used empirical force field and relied on unconfirmed assumptions regarding bath spectral densities and the level of system-specific disorder.
\textit{Ab initio} level calculations with sufficient statistics were consistently conducted to obtain reliable bath spectral densities characterizing the molecule-environmental response following the excitation of BChls and the statistics of excitation energy distributions of BChls for each of the three complexes. The efficiency of these approaches will also make them easily applicable to other LHCs toward addressing similar questions.

Overall, the bath spectral densities and reorganization energies remain quite insensitive to different sizes of complexes (see Figure~\ref{fig:spectral_density} and Table~\ref{tbl:reorg}).  In particular, the sharp peaks of the spectral density remain virtually identical, whereas the remainder part is slightly smaller for the natural one than for synthetic ones. This is understandable considering that the sharp peaks of the bath spectral density primarily reflects the intramolecular and mostly harmonic vibrational modes of BChls, whereas the remainder part represents those of all remaining but yet localized interactions surrounding BChls.   

On the other hand, we found that the quasistatic disorder, namely slow fluctuations that have classical origin and are extended over picosecond time scales, in excitation energies of BChls for the natural LH2 complex is at least half those for synthetic LH2-like complexes (see Table~\ref{tbl:std}).  
We also found that the natural LH2 complex has the smallest site-dependent heterogeneity of the disorder.   

Similar trends were found for HB.  
For the natural LH2 complex, the probability of broken HB was $0.02$, whereas it was $0.43$ and $0.81$ for the 6- and 12-fold synthetic LH2-like complexes.
This demonstrates that BChl-$\alpha$ and -$\beta$ are effectively anchored by hydrogen bonds in the natural 9-fold LH2 complex, while pigments in altered symmetries lack a stable HB network as recently confirmed for a complex with 7-fold symmetry.~\cite{cupellini-pr156}
This stabilizing network will allow pigments to be tightly packed within their binding sites, leading to the suppression of irregularities in intermolecular interactions and the quasistatic disorder.
Thus, our results demonstrate that the key molecular architecture for effectively reducing environment-induced energetic fluctuations and disorder is the establishment of an optimized HB network, anchoring BChls within the protein environment.
Indeed, in combination with the GME-MED method,\cite{jang2014,Jang2018} we further demonstrated that the inter-LH2 exciton transfer rate of the 9-fold complex is in close agreement with experimental observations and fastest among the three complexes. 
This result clearly proves that the natural sizes of LH2 complexes correspond to the design rule with most HBs and least disorder and is in line with an earlier conclusion\cite{Jang2015}, which was partly based on unconfirmed assumptions. The design rule is in regard to the dynamic aspect of LH2 and LH2-like complexes toward efficient exciton transfer, and will not be captured by simplified geometric models\cite{cleary-pnas110}.

A primary factor in tuning and regulating exciton transfer efficiencies is the control of the disorder in the energies of excitonic states.
To this end, understanding how nature-engineered antenna complexes achieve such control through their structural organization is particularly important.
The computational data and analyses provided in this work clarify this issue in regard to the relationship  between the disorder and the optimality of sizes of natural LH2 complexes.
Future applications of the methods presented in this work to other LHCs will also offer new insights into design principles in a broader context.

\begin{acknowledgement}
This work was supported by the KIAS Individual Grant (AP100301 to K.H.C.) at the Korea Institute for Advanced Study. S.J.J. acknowledges primary support from the US Department of Energy, Office of Sciences, Office of Basic Energy Sciences (DE-SC0021413, DE-SC0026114), partial support from the National Science Foundation (CHE-1900170) during the initial stage of this research, and  support from KIAS through its KIAS Scholar program. Y.M.R. additionally acknowledges the support by the National Research Foundation of Korea (NRF) grant funded by the Korean government (MSIT) through SRC with No. RS-2020-NR049542. All the authors thank the Center for Advanced Computation in KIAS for providing computing resources.
\end{acknowledgement}

\begin{suppinfo}
Computational details of quantum chemistry calculations, simulation setup, brief introduction to the IM Method, validation of ML model, detailed expressions for exciton transfer rate, ground state partial charges of BChls, benchmark of the prediction performance of the FNN, benchmark of IM/MM against reference QM/MM results, schematic overview of the sampling  protocol, benchmark of B3LYP functional against CAM-B3LYP and dispersion-corrected functionals, line shape functions and simulated absorption line shapes of 6-, 9-, and 12-fold LH2 complexes.

\end{suppinfo}
\bibliography{lh2_bib}

\end{document}


\clearpage
\section{Supporting Text}
\subsection{Quantum chemistry calculation} 
Throughout this work, quantum chemistry calculations were performed using density functional theory (DFT) method with the B3LYP functional, which has been widely adopted as a computationally efficient and numerically stable functionals with minimal parameters. We confirmed that B3LYP provides reliable results consistent with more sophisticated approaches for our specific systems, such as range-separated functionals~\cite{takeshi2004} or dispersion corrections\cite{grimme2011} (Figure S4). Besides, our IM part does not include $\pi$ stacking nor protein-ligand interaction. Thus, inclusion of dispersion interaction at the DFT level was not deemed necessary. In fact, such interactions are actually accounted for in our IM/MM model with reasonable accuracy through the IM--MM interaction terms. Potential systematic errors associated with these functionals have been discussed and benchmarked in the literature.~\cite{Jang2015}  

\subsection{Initialization and Preparation of the Simulation}
The crystal structure of the LH2 complex (PDB ID: 2FKW) was adopted as our starting point, which naturally exhibits 9-fold symmetry.
From the crystal structure, a single-unit monomer was extracted from the 9-fold symmetric complex. 
The monomers were then cylindrically arranged to construct a LH2 complex with any desired $n$-fold symmetry. 
The constructed complex was then solvated with TIP3P water model,~\cite{Jorgensen1983} accompanied by charge-neutralizing chloride ions. 
The entire solvated complex was placed at the center of a cubic simulation box with a side length of 10 nm. 
Next, we performed QM/MM energy minimization to optimize the structure using combination of GROMACS package~\cite{Pronk2013} with Q-Chem 5.0.~\cite{Epifanovsky2021}
During the optimization process, the $\pi$ conjugation system of each BChls was treated as the QM region, using density functional theory (DFT) with the 6-31G(d,p) basis set and the B3LYP functional, and the rest of the pigment was modeled to closely replicate the Hessian matrix obtained from quantum chemical calculations.~\cite{Kim2016,Song2011}
The remaining part of the system was described using the CHARMM27 force field parameters.~\cite{MacKerell1998}

\begin{table*}[t!]
\centering
  \caption{Partial charges of BChl in the ground state. Atomic indices follow the reference.~\cite{Cho2019}}
  \label{tbl:charges}
  \begin{tabular}{c|ccc|ccc|ccc}
    \hline
     & \multicolumn{3}{c|}{6-fold} & \multicolumn{3}{c|}{9-fold} & \multicolumn{3}{c}{12-fold} \\
    \hline
      & $\alpha$ & $\beta$ & $\gamma$ & $\alpha$ & $\beta$ & $\gamma$ & $\alpha$ & $\beta$ & $\gamma$ \\
    \hline
    1 & 1.47 & 1.47 & 1.52 & 1.47 & 1.63 & 1.41 & 1.45 & 1.60 & 1.55  \\
2 & $-$0.59 & $-$0.60 & $-$0.74 & $-$0.57 & $-$0.79 & $-$0.69 & $-$0.52 & $-$0.74 & $-$0.84  \\
3 & 0.11 & 0.16 & 0.26 & 0.15 & 0.35 & 0.20 & 0.02 & 0.26 & 0.27  \\
4 & 0.05 & 0.09 & 0.01 & $-$0.04 & 0.04 & $-$0.07 & 0.13 & 0.11 & 0.03  \\
5 & 0.09 & 0.11 & 0.07 & 0.11 & 0.08 & 0.14 & 0.08 & 0.05 & 0.09  \\
6 & 0.11 & 0.11 & 0.09 & 0.11 & 0.09 & 0.10 & 0.08 & 0.08 & 0.06  \\
7 & 0.02 & 0.04 & 0.06 & 0.03 & 0.02 & 0.05 & 0.03 & 0.02 & 0.05  \\
8 & 0.40 & 0.39 & 0.50 & 0.37 & 0.51 & 0.50 & 0.35 & 0.48 & 0.57  \\
9 & $-$0.12 & $-$0.33 & $-$0.13 & $-$0.08 & $-$0.19 & $-$0.30 & $-$0.17 & $-$0.10 & $-$0.23  \\
10 & 0.07 & 0.13 & 0.06 & 0.06 & 0.09 & 0.10 & 0.09 & 0.07 & 0.08  \\
11 & 0.06 & 0.08 & 0.05 & 0.04 & 0.06 & 0.11 & 0.06 & 0.03 & 0.08  \\
12 & $-$0.20 & $-$0.08 & $-$0.06 & $-$0.16 & $-$0.15 & 0.10 & $-$0.20 & $-$0.18 & $-$0.03  \\
13 & 0.10 & 0.11 & 0.03 & 0.09 & 0.12 & $-$0.03 & 0.08 & 0.11 & 0.02  \\
14 & 0.08 & 0.03 & 0.05 & 0.07 & 0.05 & 0.01 & 0.11 & 0.06 & 0.06  \\
15 & 0.69 & 0.63 & 0.66 & 0.67 & 0.66 & 0.65 & 0.70 & 0.65 & 0.63  \\
16 & $-$0.57 & $-$0.50 & $-$0.55 & $-$0.54 & $-$0.51 & $-$0.57 & $-$0.56 & $-$0.50 & $-$0.51  \\
17 & $-$0.58 & $-$0.60 & $-$0.59 & $-$0.60 & $-$0.61 & $-$0.59 & $-$0.61 & $-$0.61 & $-$0.61  \\
18 & 0.01 & 0.02 & $-$0.02 & 0.03 & $-$0.17 & 0.10 & 0.11 & $-$0.17 & 0.07  \\
19 & $-$0.31 & $-$0.40 & $-$0.41 & $-$0.33 & $-$0.37 & $-$0.45 & $-$0.30 & $-$0.37 & $-$0.42  \\
20 & 0.07 & 0.10 & 0.11 & 0.09 & 0.09 & 0.12 & 0.07 & 0.09 & 0.12  \\
21 & 0.08 & 0.11 & 0.10 & 0.09 & 0.09 & 0.12 & 0.09 & 0.09 & 0.10  \\
22 & 0.09 & 0.09 & 0.11 & 0.09 & 0.09 & 0.12 & 0.07 & 0.09 & 0.11  \\
23 & $-$0.75 & $-$0.79 & $-$0.86 & $-$0.74 & $-$0.99 & $-$0.74 & $-$0.74 & $-$0.96 & $-$0.84  \\
24 & 0.39 & 0.46 & 0.45 & 0.39 & 0.57 & 0.33 & 0.39 & 0.52 & 0.38  \\
25 & 0.16 & 0.15 & 0.20 & 0.15 & 0.12 & 0.20 & 0.15 & 0.17 & 0.21  \\
26 & $-$0.50 & $-$0.53 & $-$0.60 & $-$0.47 & $-$0.52 & $-$0.55 & $-$0.48 & $-$0.57 & $-$0.56  \\
27 & 0.51 & 0.54 & 0.66 & 0.50 & 0.68 & 0.56 & 0.50 & 0.73 & 0.64  \\
28 & 0.78 & 0.82 & 0.85 & 0.76 & 0.81 & 0.82 & 0.77 & 0.82 & 0.82  \\
29 & $-$0.59 & $-$0.64 & $-$0.65 & $-$0.57 & $-$0.63 & $-$0.65 & $-$0.56 & $-$0.59 & $-$0.63  \\
30 & $-$0.56 & $-$0.56 & $-$0.56 & $-$0.58 & $-$0.56 & $-$0.55 & $-$0.57 & $-$0.59 & $-$0.56  \\
31 & 0.15 & 0.15 & 0.15 & 0.16 & 0.15 & 0.15 & 0.15 & 0.15 & 0.15  \\
32 & 0.15 & 0.15 & 0.15 & 0.16 & 0.15 & 0.15 & 0.15 & 0.15 & 0.15  \\
33 & 0.15 & 0.16 & 0.15 & 0.16 & 0.16 & 0.15 & 0.15 & 0.16 & 0.16  \\
34 & $-$0.59 & $-$0.62 & $-$0.62 & $-$0.54 & $-$0.67 & $-$0.54 & $-$0.55 & $-$0.65 & $-$0.57  \\
35 & 0.19 & 0.19 & 0.17 & 0.17 & 0.19 & 0.16 & 0.17 & 0.18 & 0.16  \\
36 & $-$0.43 & $-$0.44 & $-$0.46 & $-$0.41 & $-$0.43 & $-$0.43 & $-$0.41 & $-$0.44 & $-$0.44  \\
37 & 0.13 & 0.14 & 0.14 & 0.13 & 0.13 & 0.13 & 0.12 & 0.13 & 0.13  \\
38 & 0.13 & 0.14 & 0.14 & 0.13 & 0.13 & 0.13 & 0.12 & 0.12 & 0.13  \\
39 & 0.13 & 0.13 & 0.13 & 0.13 & 0.14 & 0.12 & 0.12 & 0.13 & 0.13  \\
40 & $-$0.64 & $-$0.55 & $-$0.74 & $-$0.68 & $-$0.75 & $-$0.62 & $-$0.63 & $-$0.76 & $-$0.78  \\
41 & 0.32 & 0.28 & 0.43 & 0.34 & 0.38 & 0.29 & 0.32 & 0.43 & 0.46  \\
    \hline
  \end{tabular}
\end{table*}

\begin{table*}[t!]\centering
\captionsetup{labelformat=empty}
  \begin{tabular}{c|ccc|ccc|ccc}
    \hline
     & \multicolumn{3}{c|}{6-fold} & \multicolumn{3}{c|}{9-fold} & \multicolumn{3}{c}{12-fold} \\
    \hline
      & $\alpha$ & $\beta$ & $\gamma$ & $\alpha$ & $\beta$ & $\gamma$ & $\alpha$ & $\beta$ & $\gamma$ \\
    \hline
42 & 0.22 & 0.21 & 0.16 & 0.24 & 0.18 & 0.32 & 0.22 & 0.25 & 0.18  \\
43 & 0.04 & 0.05 & 0.01 & 0.05 & 0.01 & $-$0.02 & 0.02 & $-$0.02 & 0.01  \\
44 & $-$0.23 & $-$0.15 & $-$0.20 & $-$0.26 & $-$0.11 & $-$0.27 & $-$0.22 & $-$0.22 & $-$0.21  \\
45 & 0.07 & 0.08 & 0.07 & 0.07 & 0.07 & 0.10 & 0.06 & 0.10 & 0.07  \\
46 & 0.52 & 0.41 & 0.55 & 0.56 & 0.51 & 0.42 & 0.49 & 0.57 & 0.57  \\
47 & 0.09 & 0.06 & 0.19 & 0.12 & 0.06 & 0.15 & 0.14 & 0.11 & 0.19  \\
48 & $-$0.00 & 0.01 & $-$0.01 & $-$0.02 & $-$0.01 & 0.01 & $-$0.01 & $-$0.02 & 0.01  \\
49 & 0.01 & 0.01 & $-$0.03 & 0.01 & 0.00 & $-$0.02 & 0.01 & 0.00 & $-$0.03  \\
50 & $-$0.30 & $-$0.30 & $-$0.34 & $-$0.29 & $-$0.30 & $-$0.30 & $-$0.31 & $-$0.32 & $-$0.33  \\
51 & 0.08 & 0.09 & 0.09 & 0.06 & 0.09 & 0.08 & 0.06 & 0.09 & 0.09  \\
52 & 0.07 & 0.08 & 0.08 & 0.08 & 0.07 & 0.08 & 0.08 & 0.09 & 0.08  \\
53 & 0.08 & 0.09 & 0.08 & 0.07 & 0.11 & 0.08 & 0.07 & 0.10 & 0.08  \\
54 & $-$0.55 & $-$0.55 & $-$0.66 & $-$0.57 & $-$0.59 & $-$0.58 & $-$0.52 & $-$0.65 & $-$0.67  \\
55 & 0.18 & 0.19 & 0.21 & 0.17 & 0.18 & 0.20 & 0.18 & 0.21 & 0.21  \\
56 & $-$0.44 & $-$0.42 & $-$0.34 & $-$0.47 & $-$0.33 & $-$0.36 & $-$0.43 & $-$0.33 & $-$0.34  \\
57 & 0.11 & 0.11 & 0.07 & 0.12 & 0.09 & 0.09 & 0.11 & 0.09 & 0.08  \\
58 & 0.11 & 0.10 & 0.09 & 0.12 & 0.08 & 0.10 & 0.11 & 0.09 & 0.10  \\
59 & 0.11 & 0.10 & 0.10 & 0.12 & 0.08 & 0.08 & 0.12 & 0.08 & 0.09  \\
60 & $-$0.74 & $-$0.80 & $-$0.88 & $-$0.76 & $-$1.02 & $-$0.76 & $-$0.70 & $-$0.95 & $-$0.89  \\
61 & 0.42 & 0.50 & 0.52 & 0.47 & 0.66 & 0.41 & 0.44 & 0.56 & 0.55  \\
62 & 0.12 & 0.08 & 0.07 & 0.10 & 0.02 & 0.12 & 0.10 & 0.06 & 0.06  \\
63 & $-$0.34 & $-$0.30 & $-$0.34 & $-$0.33 & $-$0.31 & $-$0.32 & $-$0.30 & $-$0.33 & $-$0.34  \\
64 & 0.29 & 0.30 & 0.37 & 0.29 & 0.48 & 0.27 & 0.22 & 0.47 & 0.35  \\
65 & 0.69 & 0.73 & 0.72 & 0.71 & 0.72 & 0.68 & 0.68 & 0.71 & 0.79  \\
66 & $-$0.53 & $-$0.55 & $-$0.51 & $-$0.55 & $-$0.55 & $-$0.49 & $-$0.54 & $-$0.55 & $-$0.55  \\
67 & $-$0.64 & $-$0.74 & $-$0.69 & $-$0.71 & $-$0.64 & $-$0.67 & $-$0.67 & $-$0.63 & $-$0.86  \\
68 & 0.21 & 0.22 & 0.24 & 0.23 & 0.20 & 0.22 & 0.19 & 0.20 & 0.25  \\
69 & $-$0.10 & $-$0.18 & $-$0.10 & $-$0.08 & $-$0.10 & $-$0.12 & $-$0.07 & $-$0.11 & $-$0.08  \\
70 & 0.12 & 0.13 & 0.11 & 0.11 & 0.09 & 0.11 & 0.11 & 0.13 & 0.12  \\
71 & 0.12 & 0.14 & 0.10 & 0.11 & 0.10 & 0.12 & 0.10 & 0.08 & 0.09  \\
72 & 0.10 & 0.11 & 0.12 & 0.10 & 0.13 & 0.11 & 0.09 & 0.12 & 0.09  \\
73 & 0.69 & 0.73 & 0.73 & 0.72 & 0.74 & 0.71 & 0.72 & 0.78 & 0.84  \\
74 & $-$0.48 & $-$0.50 & $-$0.49 & $-$0.49 & $-$0.48 & $-$0.46 & $-$0.51 & $-$0.53 & $-$0.52  \\
75 & $-$0.36 & $-$0.34 & $-$0.39 & $-$0.37 & $-$0.37 & $-$0.36 & $-$0.38 & $-$0.39 & $-$0.41  \\
76 & $-$0.62 & $-$0.65 & $-$0.64 & $-$0.68 & $-$0.69 & $-$0.51 & $-$0.63 & $-$0.68 & $-$0.66  \\
77 & 0.21 & 0.20 & 0.19 & 0.21 & 0.19 & 0.18 & 0.21 & 0.20 & 0.19  \\
78 & $-$0.43 & $-$0.44 & $-$0.41 & $-$0.43 & $-$0.43 & $-$0.43 & $-$0.44 & $-$0.44 & $-$0.41  \\
79 & 0.13 & 0.14 & 0.13 & 0.13 & 0.14 & 0.14 & 0.14 & 0.14 & 0.13  \\
80 & 0.13 & 0.14 & 0.13 & 0.14 & 0.14 & 0.13 & 0.14 & 0.14 & 0.13  \\
81 & 0.13 & 0.14 & 0.13 & 0.14 & 0.14 & 0.14 & 0.14 & 0.15 & 0.13  \\
\hline
  \end{tabular}
\end{table*}

\subsection{Summary of the IM method} 
In this section, we provide a  brief overview of the PES-IM approach and relevant notations for completeness of our presentation.
This will also provide a clear motivation for employing a pre-calculated dataset to approximate QM/MM calculations.
Here, we denote the Cartesian coordinates as $\textbf{X}$ and the internal coordinates as $\textbf{Z}$.
The data point comprising the IM PES for each index $i$ consists of coordinate $\textbf{X}_i$ and relevant properties such as energy $E_i$, gradient $\textbf{g}_i$, and Hessian $\textbf{h}_i$. 
To obtain the potential energy at a pigment geometry $\textbf{Z}$, the energy is primitively approximated from each data point in the dataset using a single-point Taylor expansion:
\begin{align}
\label{eqn:im}
    V_i(\textbf{Z})=E_i + \textbf{D}^\textrm{T}_i\cdot \textbf{g}_i+\frac{1}{2}\textbf{D}^\textrm{T}_i\cdot \textbf{h}_i\cdot\textbf{D}_i
\end{align}
where $\textbf{D}_i=\textbf{Z}-\textbf{Z}_i$ represents the difference between the target geometry and the data point geometry $\textbf{Z}_i$.
The final potential energy is obtained as a weighted sum of the Taylor expansion from all data points:
\begin{align}
    V(\textbf{X})=\sum_i w_i(\textbf{X})V_i(\textbf{Z})
\end{align}
where $w_i(\textbf{X})$ is a normalized weight based on a modified Shepard weighting function.~\cite{Bettens1999}
For LH2, the $\pi$ conjugation system of every BChl was treated as the IM region.
A detailed explanation, particularly its integration with MM methods for IM/MM simulation can be found in the literature.~\cite{Kim2016}

The IM approach requires a well-designed and robust sampling scheme for collecting its
data points to ensure the accuracy and reliability of the PES. Conventionally, a primitive
database is constructed with minimally available data points, and the database is iteratively
improved by adding more data points after performing preliminary simulations and sampling
key configurations, often referred to as ``GROW scheme''.

\subsection{Inter-LH2 exciton transfer dynamics}

With a generalized master equation for modular exciton density (GME-MED), line shape functions which describe the influence of the dynamic disorder are computed as defined below:
\begin{align}
    \lambda_n &= \int^\infty_0 d \omega \frac{J_n(\omega)}{\omega}\\
    G_{n,i}(t) &= \frac{1}{\hbar} \int^\infty_0 d\omega \frac{J_n(\omega)}{\omega^2}\sin(\omega t)\\
    G_{n,r}(t) &= \frac{1}{\hbar} \int^\infty_0 d\omega \frac{J_n(\omega)}{\omega^2}\coth\left(\frac{\hbar\omega}{2k_BT}\right)\left(1-\cos(\omega t)\right)
\end{align}
where $J_n(\omega)$ denotes the spectral density of harmonic contribution for each type of BChl, as presented in the main text (Figure 4).
Figure~\ref{fig:line} shows the corresponding line shape functions for the B850 BChls of each symmetry.

The GME-MED method employs an approximation neglecting off-diagonal components in the exciton basis. Toward this, we introduce a transformation matrix between site basis of $k^{\textrm{th}}$ LH2, $\left|n_k\right\rangle$, to the exciton basis $\left|\phi_{p_k}\right\rangle$, $U_{n_k,p_k}=\left\langle n_k\right|\left.\phi_{p_k}\right\rangle$.
We also introduce reorganization energy and line shape functions in an exciton basis.
\begin{align}
    \lambda_{p_k}&=\left(\sum_{n_k} \left|U_{n_k,p_k}\right|^4\lambda_{s_n}\right) \\
    G_{p_k,i}(t)&=\left(\sum_{n_k} \left|U_{n_k,p_k}\right|^4 G_{n,i}(t)\right) \\
    G_{p_k,r}(t)&=\left(\sum_{n_k} \left|U_{n_k,p_k}\right|^4 G_{n,r}(t)\right) 
\end{align}
Then, the energy transfer rate can be approximated as
\begin{align}
    W_{1\rightarrow 2}(t)
    &=\frac{2}{\hbar^2}\textrm{Re}\sum_{p_1}\sum_{p_2}\frac{e^{-\tilde{\epsilon}_{p_1}/k_B T}}{\left(\sum_{p^\prime_1}e^{-\tilde{\epsilon}_{p^\prime_1}/k_B T}\right)}\left|\tilde{V}_{p_1 p_2}\right|^2\nonumber\\
    &\times \int^t_0 d\tau e^{-G_{p_2,r}(\tau)-i G_{p_2,i}(\tau)-i\tilde{\epsilon}_{p_2}\tau/\hbar}\nonumber\\    
    &\times \int^t_0 d\tau e^{-G_{p_1,r}(\tau)-i G_{p_1,i}(\tau)+i\tilde{\epsilon}_{p^\prime_1}\tau/\hbar}
\end{align}
with $\tilde{\epsilon}_{p_k}=\epsilon_{p_k}-\lambda_{p_k}$ and $\tilde{V}_{p_1 p_2}=\sum_{n,m} U_{n_2,p_2}V_{n_1,m_2}U^*_{m_1,p_1}$.

At the initial condition, the exciton is in thermal equilibrium in the first LH2. 
The exciton dynamics between two B850 rings follows
\begin{align}
    \frac{\partial}{\partial t} p_1(t)=W_{2\rightarrow 1}(t) p_2(t) - W_{1\rightarrow 2}(t)p_1(t)
\end{align}
With an exciton partition function of the $k$th LH2, $Z_k=\sum_{p_k} e^{-\tilde{\epsilon}_{p_k}/k_B T}$, an effective forward rate is defined as
\begin{align}
    k_f=\frac{Z_2}{Z_1+Z_2}\frac{1}{\tau_1}
\end{align}
where a transfer time $\tau_1$ is the shortest time satisfying
\begin{align}
    \ln\left( p_1(\tau_1)-\frac{Z_1}{Z_2}p_2(\tau_1) \right)=-1
\end{align}


\clearpage
\section{Supporting Figures}

\begin{figure}[h!]
\centering
\includegraphics[width=3in]{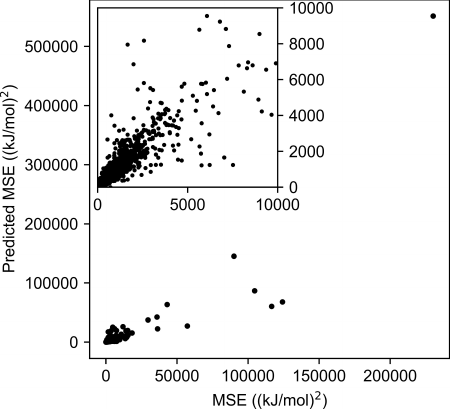}
  \caption{Correlations between the reference and the predicted values of the MSE for the test set.
  }
  \label{fig:ml_test}
\end{figure}

\begin{figure}[h!]
\centering
\includegraphics[width=3in]{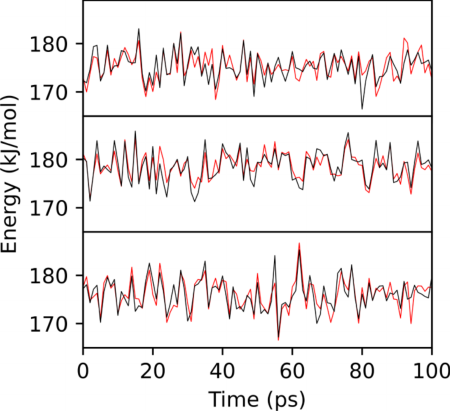}
  \caption{Time profiles of the excitation energies of BChl $\alpha$, calculated using IM/MM (red) and reference QM/MM results (black) for complexes with 6-fold (top), 9-fold (middle) and 12-fold (bottom) symmetries. Excitation energies from IM/MM are vertically shifted to align with QM/MM results.
  }
  \label{fig:funtional}
\end{figure}

\begin{figure}[]
\centering
\includegraphics[width=3in]{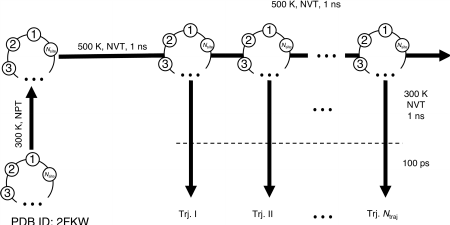}
  \caption{ A schematic overview of the sampling protocol. The procedures began with an energy-minimized structure, followed by a 1 ns equilibration in the NPT ensemble. The system was then subjected to a 2 ns annealing process at 500 K using the NVT ensemble. From the final nanosecond, ten snapshots were selected and each was equilibrated at 300 K for 1 ns under NVT conditions. Each trajectory was subsequently extended for an additional 100 ps, which constituted the production run.}
  \label{fig:mdscheme}
\end{figure}

\begin{figure}[h!]
\centering
\includegraphics[width=3in]{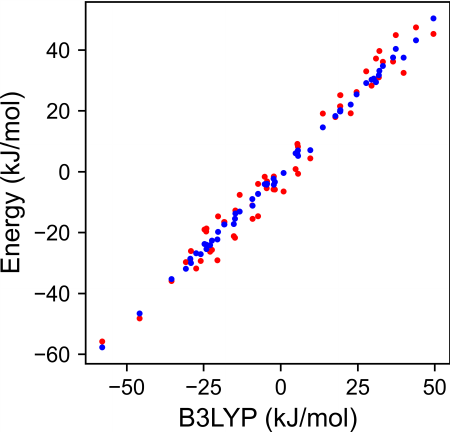}
  \caption{The agreement between B3LYP and two additional DFT methods. The data displays the energies at some trajectory-sampled geometries. The horizontal axis denotes the B3LYP energies, while the vertical axis denotes the energy from the range-separated CAM-B3LYP functional (red) or from the dispersion corrected B3LYP (blue). Because the energies in one method can be shifted by a constant, the energy at an arbitrarily selected geometry was set to zero with all the three DFT methods.}
  \label{fig:funtional}
\end{figure}

\begin{figure}[h!]
\centering
\includegraphics[width=3in]{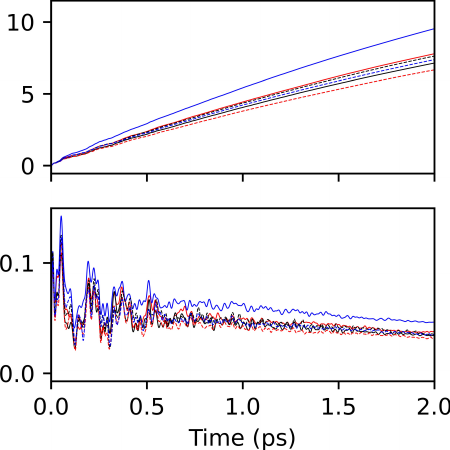}
  \caption{Real (top) and imaginary (bottom) parts of the line shape functions. Solid and dashed lines represent BChl $\alpha$ and $\beta$, while red, black, and blue correspond to 6-fold, 9-fold, and 12-fold symmetries.}
  \label{fig:line}
\end{figure}

\begin{figure}[h!]
\centering
\includegraphics[width=3in]{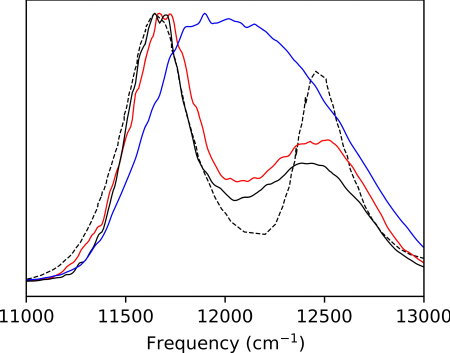}
  \caption{Calculated linear absorption line shape of LH2 complexes with 6-fold (red), 9-fold (black) and 12-fold (blue) symmetries, in comparison with experimental result (dashed).
  Total standard deviations of excitation energies for the 6-fold (12-fold) complex are  260, 252, 252 cm$^{-1}$ (269, 246, 260 cm$^{-1}$) for BChl $\alpha$, $\beta$ and $\gamma$, respectively.  
  }
  \label{fig:abs}
\end{figure}

\clearpage
\bibliography{lh2_bib}